\documentclass[aps,prb,twocolumn,superscriptaddress]{revtex4}
\usepackage{amsmath}
\usepackage{graphicx}
\usepackage{amsfonts}
\usepackage{bbold}
\usepackage{color}
\usepackage{epstopdf}

\begin{document}

\title{Chiral Y-junction of Luttinger liquid wires at weak coupling: lines
of stable fixed points}
\author{D.N. Aristov}
\affiliation{``PNPI'' NRC ``Kurchatov Institute'', Gatchina 188300, Russia}
\affiliation{Institut for Nanotechnology, Karlsruhe Institute of Technology, 76021
Karlsruhe, Germany }
\affiliation{Department of Physics, St.Petersburg State University, Ulianovskaya 1,
St.Petersburg 198504, Russia}
\author{P. W\"olfle}
\affiliation{Institut for Nanotechnology, Karlsruhe Institute of Technology, 76021
Karlsruhe, Germany }
\affiliation{Institute for Condensed Matter Theory, and Center for Functional
Nanostructures, Karlsruhe Institute of Technology, 76128 Karlsruhe, Germany}
\date{\today}

\begin{abstract}
We calculate the conductances of a Y-junction set-up of Luttinger liquid
wires threaded by a magnetic flux, allowing for different interaction
strength $g_3 \neq g$ in the third wire. The scattering matrix and the
matrix of conductances are parametrized by three variables. For these we
derive coupled RG equations, up to second order in the interaction, within
the scattering state formalism. The fixed point structure of these equations
is analyzed in detail. For repulsive interaction ($g,g_3>0$) there is only
one stable fixed point, corresponding to the complete separation of the
wires. For attractive interaction ($g<0$ and/or $g_{3}<0$) four fixed points
are found, whose stability depends on the interaction strength. For special
values of the interaction parameters (a) $g=0$, $g_{3}<0$ or (b) $%
g_{3}+g^{2}/2=0$ and $g<0$ we find whole lines of stable fixed points. We
conjecture that lines of fixed points appear generically at interfaces in
interaction coupling constant space separating two regions with different
stable fixed points. We also find that in certain regions of the $g$--$g_{3}$%
-plane the RG flow is towards a fixed point without chirality, implying that
the effect of the magnetic flux is completely screened.
\end{abstract}

\maketitle

\emph{Introduction}. Electron transport in strictly onedimensional quantum
wires is governed by the Coulomb interaction between electrons, found to
destroy the Landau quasiparticle concept valid in higher dimensions. An
important part of that physics is captured by the exactly solvable
Tomonaga-Luttinger liquid (TLL) model, in which backward scattering is
neglected.\cite{GiamarchiBook} As is well known, transport through junctions
of TLL wires is strongly affected by the interaction in the wires, sometimes
leading to a complete blocking of transmission. \ Many of the early works on
the problem of two TLL-wires connected by a junction employed the method of
bosonization \cite{Kane1992,Furusaki1993}, or, for special values of the
interaction parameter, used a mapping on to exactly solvable models\cite%
{Weiss1995,Fendley1995}. An alternative formulation using the fermionic
representation has been pioneered by \cite{Yue1994}. There the
renormalization group (RG) method has been used. The advantage of the latter
formulation is that it is rather flexible, allowing for the inclusion of
spin, of backward scattering, the study of wires with two impurities \cite%
{Polyakov2003}, and of multi-lead junctions\cite{Das2004,Lal2002} in and out
of equilibrium. The limitation of the original model calculation \cite%
{Yue1994} to weak coupling has prevented a more widely spread use of this
method. It may be shown, however, that the method may be extended to apply
in the strong coupling regime as well, by summing up an infinite class of
leading contributions in perturbation theory\cite{Aristov2009}. Quite
generally, the transport behavior at low energy/temperature is dominated by
only a few fixed points of the RG flow. The most common and intuitively
plausible fixed points are those with quantized conduction values, $G=0,1$,
in units of the conductance quantum $G_{0}=e^{2}/h$. There may, however,
appear additional fixed points associated with noninteger conductance
values. A particularly interesting case is that of a symmetric three-lead
junction with broken chiral symmetry, as induced by a magnetic flux
threading the junction. This latter problem has been studied by the method
of bosonization in \cite{Chamon2003,Oshikawa2006}, and by the functional RG
method in\cite{Barnabe2005}. These authors have identified a number of new
fixed points, in particular for attractive interaction. The discovery of
these new fixed points has sparked interest in the problem of mapping out
the complete fixed point structure of the theory, even though most of the
interesting new physics appears to be in the physically less accessible
domain of strongly attractive interactions. Only the fully symmetric
situation of identical wires ($g=g_{3}$, see below) has been considered in 
\cite{Oshikawa2006,Barnabe2005}. As we show below, by restricting the
consideration to the symmetric case, important new physics is missed. We
mention in passing that a very general study of junctions of $n$ quantum
wires using bosonization and scale invariance can be found in\cite%
{Bellazzini2007}. The effect of spin on chiral Y-junctions has been
considered very recently in\cite{Hou2008,Hou2009}.

Experimentally the Y-junction set-up may be realized by a one-dimensional
tunneling tip contacting a quantum wire. Although Y-junctions threaded by
magnetic flux have not yet been studied in experiment (the flux generated at
a nanoscale junction by an externally applied magnetic field will be too
small, but local magnetic moments generating internal flux may be an
option), similar set-ups with either broken time reversal symmetry or broken
parity have been considered recently. The tunneling of spin-polarized
electrons into the edge channel of a spin Hall effect system has been
theoretically studied in\cite{Das2011}. The asymmetry of the injected
current calculated there is reminiscent of the \textquotedblleft Hall
current\textquotedblright\ (conductivity $G_{ab}$) studied below. However,
the effect considered in\cite{Das2011} involves the symmetric component of
the off-diagonal conductivity (in our notation $G_{ab}=G_{ba}$). Similarly,
by injecting current into the edge states of a semiconductor Hall bar in the
presence of a suitable magnetic field, an asymmetry of the currents at the
two half wires has been detected, depending on the interaction strength in
the wire\cite{Steinberg2008}. The latter has been interpreted as a signature
of charge fractionalization.

In this paper we report results of an RG treatment of electron transport in
the linear response regime through a junction of three spinless TLL wires
threaded by a magnetic field. We employ a fermionic representation as
described in detail in\cite{Aristov2009}. A comparison of this method in its
simplest form with the functional renormalization group method has been made
recently in\cite{Meden2008}. The method has formerly been used in \cite%
{Das2004,Lal2002} in the case of Y-junctions. More recently this approach
has been used to derive the RG equations for the conductances of a
Y-junction connecting three TLL wires in the absence of magnetic flux, for
weak interaction \cite{Aristov2010}, and in a recent work for any strength
of interaction \cite{Aristov2011a}. In\cite{Aristov2010} it was found that
even in weak coupling, but beyond lowest order, interesting new structures
appear. The most important result there is that the fixed point found in
lowest order for repulsive interaction, describing the separation of the
third wire (the tunneling tip) from the ideally conducting main wire, is
actually a saddle point and is thus unstable. The message here is that even
a weak higher order contribution may change the RG-flow in a dramatic way.
In the present paper we describe a similar effect: an asymmetry of the three
wires of a chiral Y-junction to the effect that in the tip wire the
interaction strength $g_{3}$ is different from that in the main wire, $g$,
allows to access certain lines in interaction parameter space where a whole
line of fixed points rather than a single fixed point is stable. This
happens for attractive interaction only ($g,g_{3}<0$) \ The line of stable
fixed points is connecting two fixed points at two special manifolds of
interaction values, (a) $g=0$; $g_{3}<0$, or (b) whenever the condition $%
g_{3}=-g^{2}/2$ is met. Along these fixed point lines the FP values of the
conductances are continuously varying. To our knowledge this is the first
time that lines of stable fixed points of the conductance have been found in
models of TLL-wires. \cite{footnote} 

\begin{figure}[tb]
\includegraphics*[width=0.75\columnwidth]{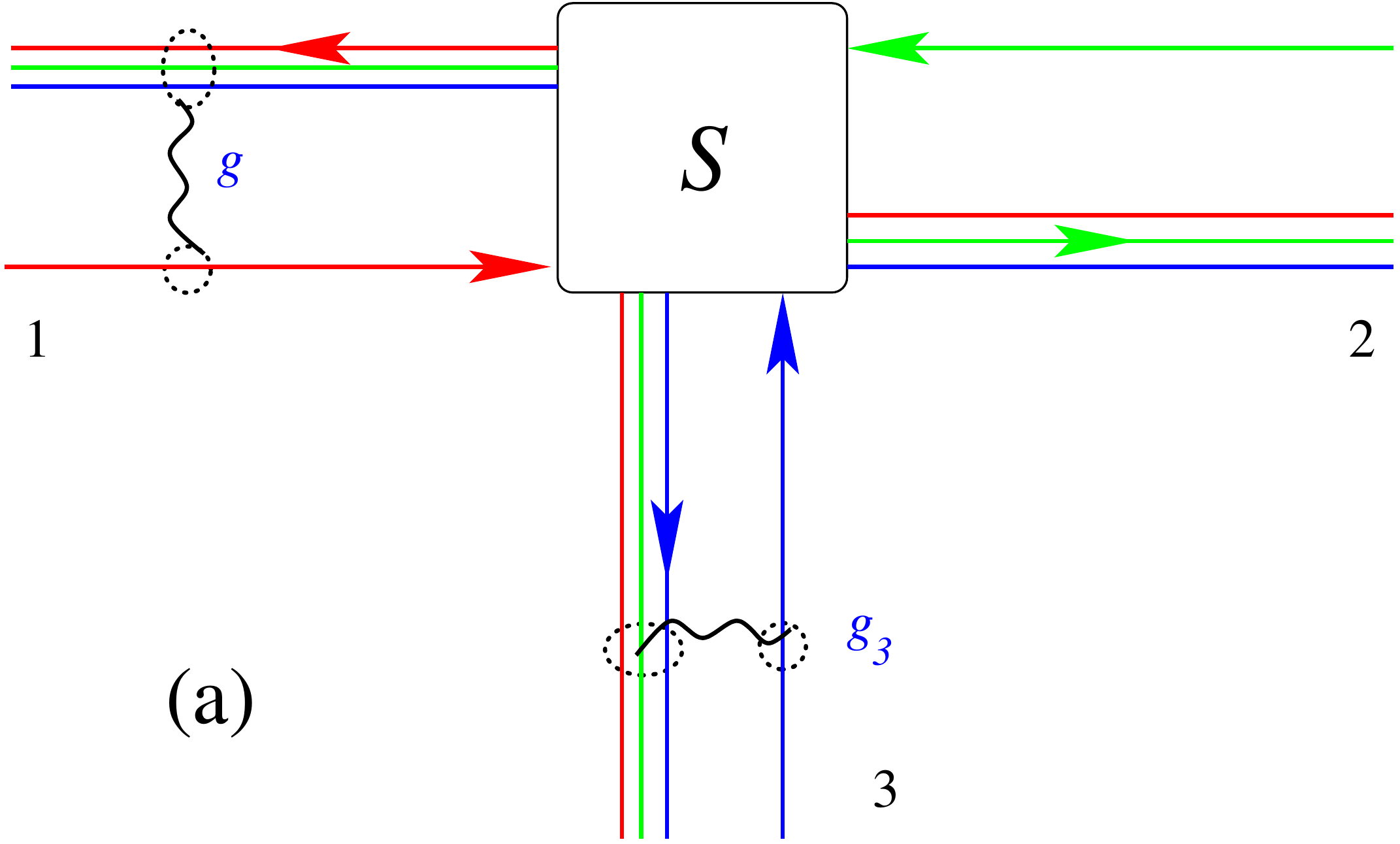}
\includegraphics*[width=0.8\columnwidth]{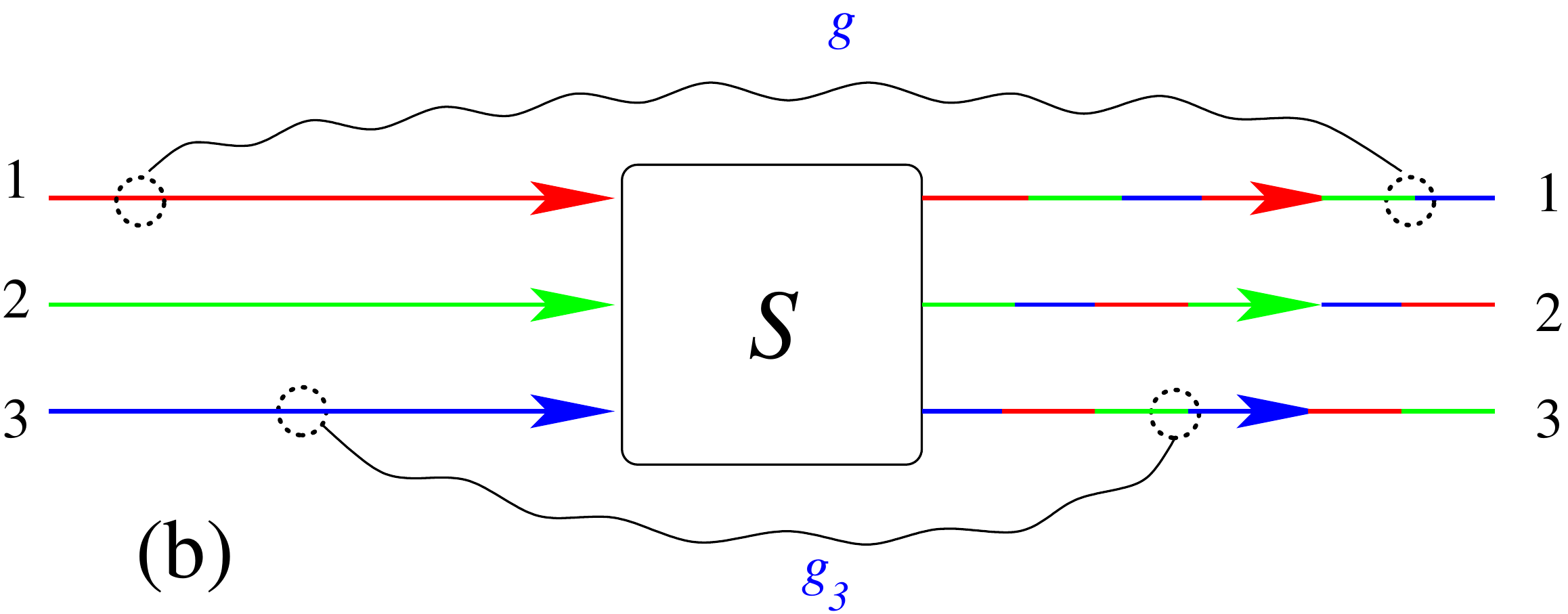}
\caption{(Color online)  (a) The geometry of a $Y$-junction is shown together
with currents of incoming and outgoing fermions, the type of fermions in the
scattered states representation is indicated by its color. The 
short-range interaction between the fermions is shown by wavy lines. (b)
The equivalent representation in the chiral fermion basis after the
unfolding procedure; the initially local interaction becomes non-local.
}
\label{fig:junction}
\end{figure}

\emph{The model}. We consider a model of interacting spin-less fermions
describing three quantum wires connected at a single junction by tunneling
amplitudes in the presence of a magnetic flux $\phi =\Phi /\Phi _{0}$
piercing the junction (with $\Phi _{0}$ the flux quantum). In the continuum
limit, linearizing the spectrum at the Fermi energy and including forward
scattering interaction of strength $g_{j}$ in wire $j$ , we may write the
TLL Hamiltonian in the representation of incoming and outgoing waves as 
\begin{eqnarray} 
\mathcal{H} &=& \int_{-\infty}^{0}dx\sum_{j=1}^{3}[H_{j}^{0}+H_{j}^{int}]\,,
\nonumber \\
H_{j}^{0}&=&v_{F}\psi_{j,in}^{\dagger
}i\nabla\psi_{j,in}-v_{F}\psi_{j,out}^{\dagger }i\nabla\psi_{j,out}\,,\\
H_{j}^{int}&=&2\pi v_{F}g_{j}\psi_{j,in}^{\dagger
}\psi_{j,in}\psi_{j,out}^{\dagger }\psi_{j,out} \, \Theta(x;-L,-l)\,. 
\nonumber \end{eqnarray} 
Here $\Theta(x;-L,-l)=1$, if $-L<x<-l$ and zero otherwise, where $L$ is the length 
of the interacting region in each wire and $l$ is the size of the junction,
inside which interaction is assumed to be absent. 
The regions $x<-L$  are thus regarded as reservoirs or leads labeled $j=1,2,3$.  
We assume that the boundaries at $x=-L$ do not produce additional potential scattering. 
The so-called backward scattering in the wires is 
unimportant in the considered spin-less situation, as it can be included 
into the $g_{j}$ amplitudes. \cite{GiamarchiBook}
We denote $g_{1}=g_{2}=g$ from now on, and put $%
v_{F}=1$. The various incoming and outgoing channels are illustrated in Fig.\ \ref{fig:junction}. 

In terms of the triplet of incoming fermions $\Psi =(\psi
_{1,in},\psi _{2,in},\psi _{3,in})$ the outgoing fermion operators may be
expressed as $\Psi (x)=S\cdot \Psi (-x)\,$. Here the 3$\times $3 $S$-matrix
characterizes the scattering at the junction and (up to irrelevant phase
factors) has the structure 
\begin{equation}
\begin{aligned} S&=\begin{pmatrix}r_{1},&t_{12},&t_{13}\\
t_{21},&r_{1},&t_{23}\\ t_{31}&t_{32}&r_{2}\end{pmatrix}\,,\\
r_{1}&=\tfrac{1}{2}\left(\cos\theta+e^{i\psi}\right)\sin\xi,\quad
r_{2}=\cos\theta \,,\\ t_{13}&=t_{32}=i\cos\tfrac{\xi}{2}\sin\theta,\quad
t_{23}=t_{31}=i\sin\theta\sin\tfrac{\xi}{2}\,,\\
t_{12}&=\cos\theta\cos^2\tfrac{\xi}{2}-e^{i\psi} \sin^2\tfrac{\xi}{2},\quad
t_{21}=t_{12}\vert_{\xi\to\pi-\xi}\,. \end{aligned}  \label{Smatrix}
\end{equation}

We express the interaction term of the Hamiltonian in terms of density
operators $\widehat{\rho }_{j,in}=\psi _{j,in}^{\dagger }\psi _{j,in}=\Psi
^{+}\rho _{j}\Psi =\widehat{\rho }_{j}$, and $\widehat{\rho }_{j,out}=\psi
_{j,out}^{\dagger }\psi _{j,out}=\Psi ^{+}\widetilde{\rho }_{j}\Psi =%
\widehat{\widetilde{\rho }}_{j}$, where $\widetilde{\rho }_{j}=S^{+}\cdot
\rho _{j}\cdot S$, as 
\begin{equation}
H_{j}^{int}=2\pi \{g[\widehat{\rho }_{1}\widehat{\widetilde{\rho }}_{1}+%
\widehat{\rho }_{2}\widehat{\widetilde{\rho }}_{2}]+g_{3}\widehat{\rho }_{3}%
\widehat{\widetilde{\rho }}_{3}\} \Theta(x;-L,-l)\,.
\end{equation}%
The density matrices are given by $(\rho _{j})_{\alpha \beta }=\delta
_{\alpha \beta }\delta _{\alpha j}$ and $(\widetilde{\rho }_{j})_{\alpha
\beta }=S_{\alpha j}^{+}S_{j\beta }$. A convenient representation of 3$%
\times $3-matrices is in terms of Gell-Mann matrices $\lambda _{j}$, $%
j=0,1,\ldots ,8$, the generators of $SU(3)$ (see\cite{Aristov2011a}). Notice
that the interaction operator only involves $\lambda _{3}$ and $\lambda _{8}$
(besides the unit operator $(\lambda _{0})_{\alpha \beta }=\sqrt{2/3}%
\,\delta _{\alpha \beta }$). We note $Tr(\lambda _{j})=0$, $Tr(\lambda
_{j}\lambda _{k})=2\delta _{jk}$, $j=1,\ldots ,8$, and $[\lambda
_{3},\lambda _{8}]=0$. Introducing a compact notation we define a
three-component vector $\mathbf{\lambda }=(\lambda _{3},\lambda _{8},\lambda
_{0})$, in terms of which the densitries may be expressed as $\rho _{j}=%
\sqrt{1/2}\sum_{\mu }R_{j\mu }\mathbf{\lambda }_{\mu }$, where the $3\times
3 $ matrix $\mathbf{R}$ is defined as

\begin{equation}
\mathbf{R}=\left( 
\begin{array}{ccc}
\frac{1}{\sqrt{2}} & \frac{1}{\sqrt{6}} & \frac{1}{\sqrt{3}} \\ 
-\frac{1}{\sqrt{2}} & \frac{1}{\sqrt{6}} & \frac{1}{\sqrt{3}} \\ 
0 & -\sqrt{\frac{2}{3}} & \frac{1}{\sqrt{3}}%
\end{array}%
\right)
\end{equation}%
and has the properties $\mathbf{R}^{-1}=\mathbf{R}^{T}$, $det\,\mathbf{R}=1$%
. The outgoing amplitudes are expressed in the analogous form with $\lambda
_{j}$ replaced by $\widetilde{\lambda }_{j}=S^{+}\cdot \lambda _{j}\cdot S$
. With the aid of the $\lambda _{j}$ the $S$-matrix may be parametrized by
eight angular variables (see\cite{Aristov2011}). For the case under
consideration only three of these, $\theta ,\psi ,\xi $, are relevant: $%
S=e^{i\lambda _{2}\xi /2}e^{i\lambda _{3}(\pi -\psi )}e^{i\lambda _{5}\theta
}e^{i\lambda _{2}(\pi -\xi )/2}$ ; this $S$-matrix is given explicitly by
Eq.\ (\ref{Smatrix}). As will be shown below, the S-matrix and therefore the
angular variables $\theta ,\psi ,\xi $, will be renormalized by the
interaction.

\begin{figure}[tb]
\includegraphics*[width=0.7\columnwidth]{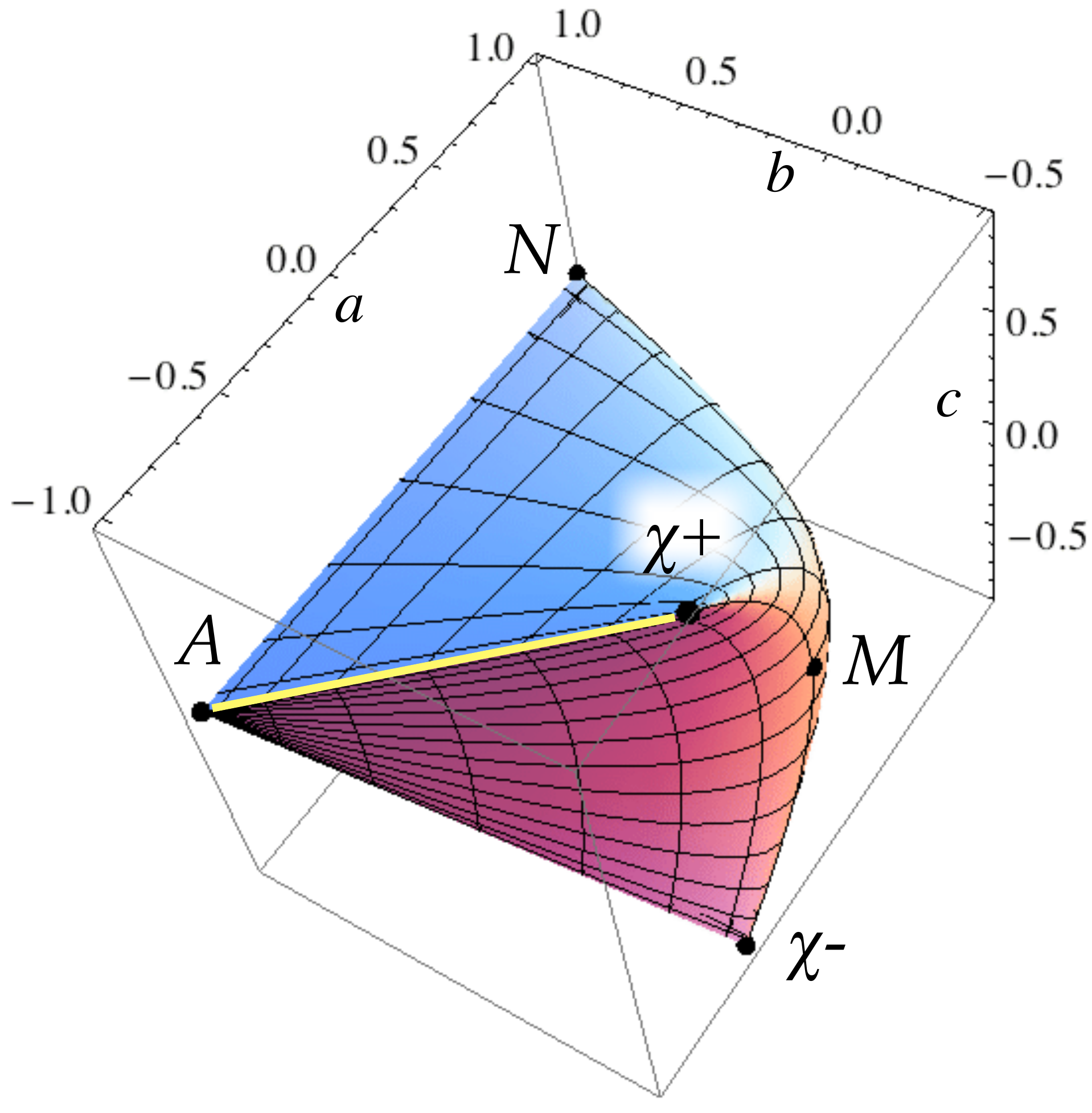}
\caption{(Color online) Allowed values of conductances, $a,b,c$, lie inside
the depicted body. The location of RG fixed points $N$, $A$, $\protect\chi%
^{\pm}$, $M$ is also shown. The line of fixed points connecting FPs $A$ and $%
\protect\chi^{+}$ is indicated in yellow. }
\label{fig:body}
\end{figure}

\emph{Parametrization of conductance matrix}. In the linear response regime,
we may define a matrix of conductances $C_{jk}$ relating the current $I_{j}$
in lead $j$ (flowing towards the junction) to the electrical potential $%
V_{k} $ in lead $k$ : $I_{j}=\sum_{k}C_{jk}V_{k}$. The conductance matrix $%
\mathbf{C}$ has only three independent elements, which may be represented in
the form of a $2\times 2$ antisymmetric matrix $\mathbf{G}$ with elements $%
G_{11}=G_{aa}=\frac{1}{2}(1-a)$, $G_{22}=G_{bb}=\frac{2}{3}(1-b)$, $%
G_{12}=-G_{21}=G_{ab}=\frac{c}{\sqrt{3}}$. These conductances relate the
currents $I_{a}=\frac{1}{2}(I_{1}-I_{2})$, $I_{b}=\frac{1}{3}%
(I_{1}+I_{2}-2I_{3})$ to the bias voltages $V_{a}=(V_{1}-V_{2})$, $V_{b}=%
\frac{1}{2}(V_{1}+V_{2}-2V_{3})$. \ It is useful to note that the nonzero
elements of the matrix $\mathbf{C}^{R}=\mathbf{R}^{T}\cdot \mathbf{C\cdot R}$
are essentially the reduced conductances: $C_{11}^{R}=2G_{aa}$ , $%
C_{22}^{R}=(3/2)G_{bb}$ and $C_{12}^{R}=-C_{21}^{R}=\sqrt{3}G_{ab}$ , where
the numerical factors arise due to the physically motivated asymmetric
definitions of the currents and voltages. We now show that the
parametrization of conductance in terms of $a$, $b$, $c$ follows quite
naturally, by observing that the initial conductances are given by $%
C_{jk}=\delta _{jk}-Tr(\widetilde{\rho }_{j}^{r}\rho _{k})=\delta
_{jk}-|S_{jk}^{r}|^{2}$, where the label $r$ indicates that the quantity is
fully renormalized by interactions. \cite{Aristov2011a}
By expressing $Y_{jk}=Tr(\widetilde{\rho 
}_{j}^{r}\rho _{k})$ in terms of $Y_{\mu \nu }^{R}=\frac{1}{2}Tr(\widetilde{%
\lambda }_{\mu }^{r}\lambda _{\nu })$ as $\mathbf{Y}=(\mathbf{R\cdot Y}%
^{R}\cdot \mathbf{R}^{T}\mathbf{)}$, we find by comparison with conductance
matrix $\mathbf{C}^{R}$ that $\mathbf{Y}^{R}$ has nonzero elements given by
the conductance parameters introduced above $Y_{11}^{R}=a$, $Y_{22}^{R}=b$, $%
Y_{12}^{R}=-Y_{21}^{R}=c$, $Y_{33}^{R}=1$. Therefore, we may use $\mathbf{Y}%
^{R}$ to represent the conductances in the renormalization group analysis
below. By substituting the $S$-matrix in the form (\ref{Smatrix}) into the
definition of $\mathbf{Y}^{R}$, we find the relation of the $a$, $b$, $c$ to
the angle variables: $a=-\frac{1}{2}(\cos ^{2}\theta +1)\cos ^{2}\xi +\cos
\theta \cos \psi \sin ^{2}\xi $, $b=\frac{1}{2}(3\cos ^{2}\theta -1)$, $c=-%
\frac{\sqrt{3}}{2}\sin ^{2}\theta \cos \xi $. We find therefore that $a,b,c$
are confined within the region $a\in \lbrack -1,1]$, $b\in \lbrack -1/2,1]$, 
$c\in \lbrack -\sqrt{3}/2,\sqrt{3}/2]$. The physically allowed points in $a$-%
$b$-$c$-space are inside a complex-shaped body labelled $B$ as shown in
Fig.\ \ref{fig:body}.

\emph{Renormalization group equations}. The renormalization of the
conductances by the interaction is determined by first calculating the
correction terms in each order of perturbation theory. We are in particular
interested in the scale-dependent contributions proportional to $%
\Lambda=\ln(L/l)$ , where $L$ and $l$ are two above lengths, characterizing the interaction region in the wires. Here we only consider terms
up to second order in $g,g_{3}$.

The conductances are given by $C_{jk}=\int_{-\infty }^{-L}dy\,\langle 
(\widehat{\rho }_{j}(-L) - \widehat{\widetilde{\rho }}_{j}(L) )\widehat{\rho }_{k}(y)\rangle _{\omega \rightarrow 0}$. Here the combination $\widehat{\rho }_{j}(-L) - \widehat{\widetilde{\rho }}_{j}(L)$ means that the current is measured at $x=-L$,  
and integration over $y$ for the \emph{incoming} densities corresponds to the notion of  voltage, $V_{k}$, applied to the leads; in the considered limit $\omega L \rightarrow 0$ the value of $L$ in this definition does not play a role.  \cite{Aristov2009} 
The quantum averaging $\langle \ldots \rangle$ is accompanied by taking trace over the matrices' product.
Up to second
order we need to consider the contributions depicted in Fig.\ 4 of Ref.\ 
\cite{Aristov2010}. Differentiating these results with respect to $\Lambda $
(and then putting $\Lambda =0$) we find the RG equations up to second order
in the interaction 
\begin{equation*}
\tfrac{d}{d\Lambda }Y_{\mu \nu }^{R}=-\tfrac{1}{2}Tr\left( \widehat{W}^{R}
_{\mu \nu }\widehat{W}^{R}_{\mu ^{\prime }\nu ^{\prime }}\right) (g_{\nu
^{\prime }\mu ^{\prime }}^{R}+\tfrac{1}{2}g_{\nu ^{\prime }\kappa
}^{R}Y_{\kappa \kappa ^{\prime }}^{R}g_{\kappa ^{\prime }\mu ^{\prime
}}^{R})\,.
\end{equation*}%
Here the $\widehat{W}^{R}_{\mu \nu }=(\mathbf{R}^{T}\cdot \widehat{\mathbf{W}%
}\cdot \mathbf{R)}_{\mu \nu }=\frac{1}{2}[\lambda _{\mu },\widetilde{\lambda 
}_{\nu }]$, and $\widehat{W}_{jk}=[\rho _{j},\widetilde{\rho }_{k}]$, are 
3$\times $3 matrices for each pair of $\mu \nu $ (or $jk$) and the trace
operation $Tr$ is defined with respect to that matrix space, whereas $g_{\mu
\nu }^{R}=(\mathbf{R}^{T}\cdot \mathbf{g\cdot R)}_{\mu \nu }$, with 
$g_{jk}=\delta _{jk}g_{j}$ and $Y_{\kappa \kappa ^{\prime }}^{R}$ are scalars
with respect to this space (sums over repeated indices are implied). The
nine matrices $\widehat{W}^{R}_{\mu\nu}$ may be evaluated best with the
aid of computer algebra, inserting the $S$-matrix in terms of the quantities 
$a,b,c$, as given above. As a result one finds the following set of
RG-equations 
\begin{eqnarray}
\tfrac{da}{d\Lambda } &=&\beta _{a}=\tfrac{g}{2}(A_{1}+\tfrac{1}{\kappa }%
A_{2}+\tfrac{g}{4}A_{3})\,,  \notag \\
\tfrac{db}{d\Lambda } &=&\beta _{b}=\tfrac{g}{2}(A_{2}+\tfrac{1}{\kappa }%
B_{2}+\tfrac{g}{4}B_{3})\,, \\
\tfrac{dc}{d\Lambda } &=&\beta _{c}=-c\tfrac{g}{2}[1+2a+\tfrac{1}{\kappa }%
(1+2b)+\tfrac{g}{4}C_{3}]\,,  \notag
\end{eqnarray}%
where we defined
\begin{equation}\begin{aligned} 
A_{1} &=1+b-2a^{2}, \quad
A_{2} = a(1-b)+c^{2}\,, \\
B_{2} &=(1-b)(1+2b), \\
A_{3} &=
 a(3+b-4 a^2)+c^2 +\tfrac{2}{\kappa } (ab -a 
   +c^2(4 a+1) )
   \\  & +\tfrac1{\kappa ^2} (1+2 b)(a
   -ab+c^2) \,, \\ 
   B_{3} &=
(1-b) \left(2 a^2+2 b+1\right) + 2 a
   c^2  \\ & +
   \tfrac2{\kappa} (2 b+1) \left(b+2 c^2-1\right) 
   +  \tfrac1{\kappa ^2}(1+3 b-4 b^3) \,, \\ 
  C_{3} &= 
 4 a^2+2 a+2 b+1 -  \tfrac2{\kappa}\left(a
   (1-b)+3 \left(b+c^2\right)\right) 
   \\ & + \tfrac1{ \kappa ^2}(2 b+1)^2 \,,
\end{aligned}\end{equation} 
and $\kappa =3g/(g+2g_{3})$.

\emph{Fixed points of RG flow}. The fixed points of the above RG equations
may be found analytically. For general (but small) values of $g,g_{3}$ we
find five fixed points in total, labelled $N,A,M,\chi ^{+},\chi ^{-}$. At
fixed point $N$ we have $a=b=1$ and $c=0$. It corresponds to the complete
separation of all wires and $G_{aa}=G_{bb}=G_{ab}=0$.  The approach to the
fixed point is given by the power laws $G_{bb}\propto L^{-\gamma
_{1}^{N}}$ \ and $4G_{aa}-G_{bb} \propto L^{-\gamma _{2}^{N}}$ with
exponents $\gamma _{1}^{N}=g+g_{3}+g^{2}/2$ and $\gamma
_{2}^{N}=2g+g^{2}$. The third direction of RG flow, 
along the $c$ axis in Fig.\ \ref{fig:body}, 
is unaccessible due to unitarity restrictions, see \cite{Aristov2011}.
This fixed point is the only stable one for
repulsive interaction, either $g>0$ or $g_{3}>0$, or both. 

Fixed point $A$
is characterized by $a=-1$, $b=1$, $c=0$ corresponding to $G_{aa}=1$, 
$G_{bb}=G_{ab}=0$ and thus the situation of a perfect wire 1-2, detached from
wire $3$. Stability analysis shows that it is stable for attractive
interaction, $g<0$, provided $g_{3}+ g^{2}/2>0$. The critical exponents
at $A$ are defined as $4(G_{aa}-1)+G_{bb}\propto L^{-\gamma _{1}^{A}}$ and 
$G_{bb}, G_{cc}\propto L^{-\gamma _{2}^{A}}$ and are
given by $\gamma_{1}^{A}=2g-g^{2}$, $\gamma _{2}^{A}=-g_{3}-g^{2}/2$. 

\begin{figure}[tb]
\includegraphics*[ width=0.7\columnwidth]{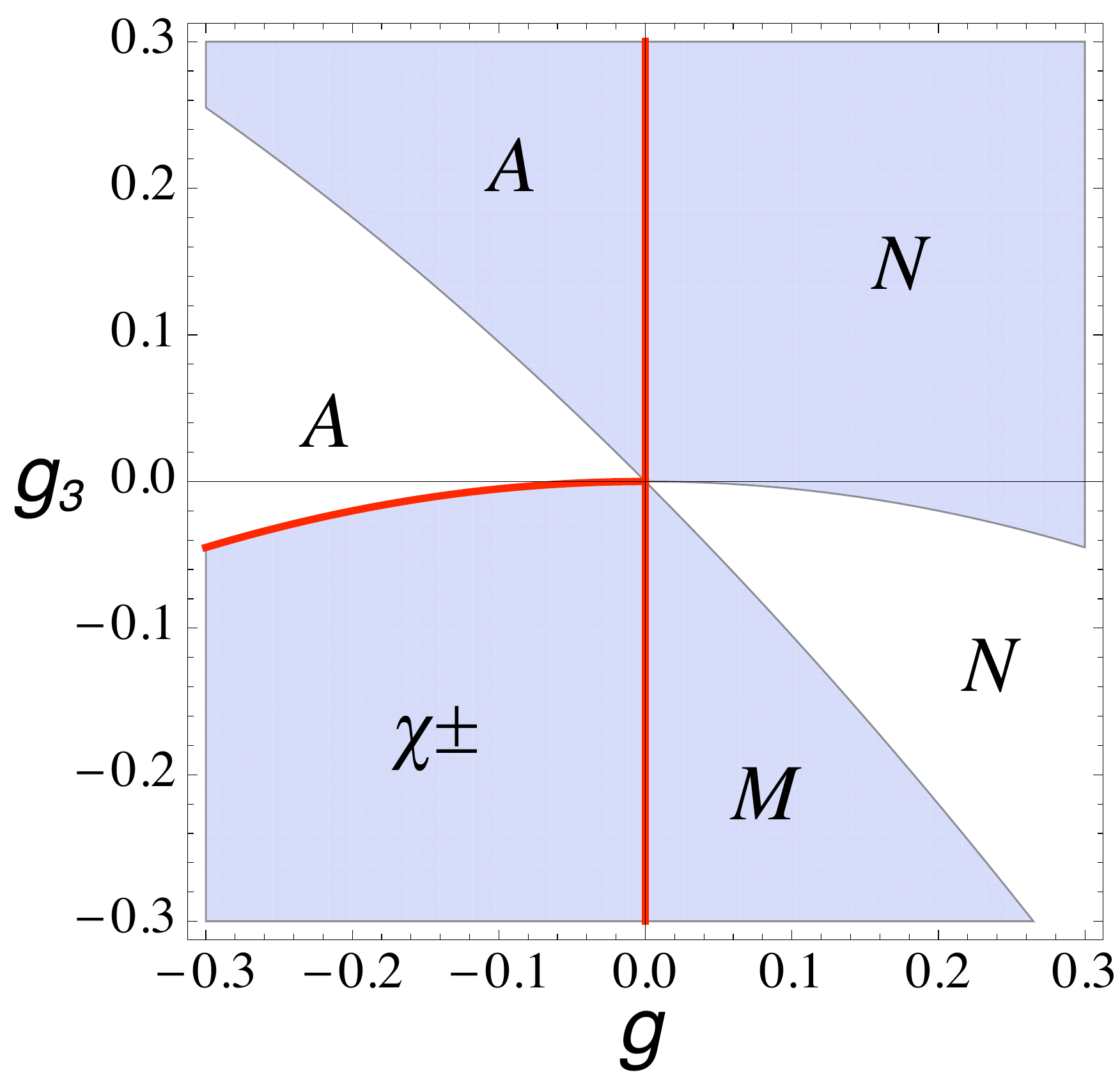}
\caption{(Color online) The RG phase portrait, showing regions of stability
of fixed points in the space of interactions $g$ and $g_{3}$. The M point
only exists in the shaded regions. Lines of fixed points appear for $g,g_{3}$
values located along the red (thick) lines. }
\label{fig:RGportrait}
\end{figure}

Fixed point $M$ is characterized by noninteger values of the conductance,
which are found to depend on the strength of interaction, $a\simeq -\kappa
/3+g h $, $b\simeq (\kappa ^{2}/3-1)/2-g\kappa h$, $c=0$, with $h=\kappa(\kappa-1)/6$. It is stable for $g>0$
and $g_{3}<0$ in a region depicted in Fig.\ \ref{fig:RGportrait}. The
critical exponents at $M$ are obtained as $\gamma_{1}^{M}\simeq g (\kappa-9\kappa^{-1})/12$,  $\gamma_{2}^{M}\simeq -g (\kappa+3\kappa^{-1})/2$ and 
$\gamma_{3}^{M}\simeq g (3-\kappa)/6$, in the scaling laws 
$\frac{4\kappa}{3}G_{aa} + G_{bb} - \big(1+\frac\kappa3\big)^{2} \propto L^{-\gamma_{1}^{M}}$, 
$ \frac{\kappa^{2}-9}{6\kappa} G_{aa} + G_{bb} +\frac{(\kappa-3)^{2}(\kappa+3)}{12\kappa} \propto L^{-\gamma_{2}^{M}}$, and $G_{cc}\propto L^{-\gamma _{3}^{M}}$.  
The region of stability of $M$ corresponds to $0<g\alt -g_{3}$, which translates to $-3\alt\kappa <0$, in which range all $\gamma_{i}^{M} >0$. 

Whereas
fixed points $N,A,M$ exist also in the absence of magnetic flux, two new (stable)
fixed points, $\chi ^{\pm }$, appear as a result of the broken time reversal
symmetry. These are the fixed points of maximum chirality first discussed in 
\cite{Chamon2003,Oshikawa2006}. The corresponding fixed point values are 
$a=b=-\frac{1}{2}$ and $c=\pm \frac{1}{2}\sqrt{3}$ , or in terms of
conductances: $G_{aa}=\frac{3}{4}$ , $G_{bb}=\frac{1}{2}$, and 
$G_{ab}=-G_{ba}=\pm \frac{1}{2}$. The fixed points $\chi ^{\pm }$ are stable
for attractive interaction, $g,g_{3}<0$, in the domain $|g_{3}|>g^{2}$ (see
Fig.\ref{fig:RGportrait}). In this case the conductances are not quantized
in integer units of the conductance but assume fractional values. The two
fixed points $\chi ^{\pm }$ are located symmetrically to the plane $c=0$ in 
$a$-$b$-$c$-space. Depending on whether the initial condition for $c$ is $%
c_{in}>0$ or $c_{in}<0$ the RG-flow will tend to $\chi ^{+}$ or $\chi ^{-}$.
If $c_{in}=0$ the flow will stay in the $c=0$ plane; the $M$ point is stable
in place of $\chi ^{\pm }$ in this case\cite{Aristov2011a}. The critical
exponents at $\chi ^{\pm }$ are found as $\gamma _{1}^{\chi}= g$ and $\gamma _{2}^{\chi}= g_{3}+g^{2}/2$. 

In the
case $g_{3}=g$ the values of the critical exponents 
at the fixed points $N$, $A$, $\chi^{\pm}$  obtained here agree
with the results quoted in the literature \cite{Chamon2003,Oshikawa2006}. 
The comparison of our results with those obtained by the bosonization method
is nontrivial, since the latter are valid for infinite TLL wires, whereas our method provides results for the experimentally relevant case of finite TLL wires attached to ideal leads. Following earlier works, Oshikawa et al.\ \cite{Oshikawa2006} have defined a prescription 
for calculating the conductances ${ C}_{ij}$ of wires attached to ideal leads from the conductances ${\bar C}_{ij}$ of the infinite wire system, see Sec.\ 12 therein. 
In the following we shall use their recipe for converting ${\bar C}_{ij}$ into ${ C}_{ij}$. 

Fixed point $M$ requires special consideration. Its existence had been first conjectured in\cite{Chamon2003,Oshikawa2006}, but its properties were not accessible.  Later fixed point $M$ was clearly identified in numerical studies using the functional renormalization group method \cite{Barnabe2005}. The above values of the critical exponents [see also \cite{Aristov2011a}], 
$\gamma_{2}^{M}=-2g \simeq 2(K-1)$ and $\gamma_{3}^{M}=g/3 \simeq (1-K)/3$,
 with $K = \sqrt{(1-g)/(1+g)}$ the Luttinger parameter, are in agreement with the numerical estimates obtained in \cite{Barnabe2005}.
 Recent numerical results \cite{Rahmani2010}
on the values of conductances ${\bar C}_{ij}$ at $\chi^{\pm}$ and $M$ obtained via the Density Matrix Renormalization Group (DMRG) method confirm the known analytical results for the $\chi^{\pm}$ fixed points, in agreement with our results. In Ref.\ \cite{Aristov2011a}  (see also above) we found the value of the conductances at the $M$ point,  $C_{ij} = \frac49 (3\delta_{ij}-1)$, independent of the strength of interaction, $g=g_{3}$. Expressing the results in terms of conductances ${\bar C}_{ij}$, we find  ${\bar C}_{ij} = C_{ij} \frac{3K}{2+K}$. The component ${\bar C}_{12}$ agrees very well with the numerical estimate given in  \cite{Rahmani2010}, lending support to our identification of the fixed point $M$ with the ``mysterious'' fixed point in \cite{Oshikawa2006}. 

It is worth noting that in the region of stability of the $M$ point the initial
Hall conductance scales to zero, meaning that the magnetic flux will be
screened (this is also true when the $N$ or $A$ points are stable, but for a
trivial reason, since at least one of the wires gets separated in that
case). In Fig.\ \ref{fig:body} we show the location of the fixed points in 
$a$-$b$-$c$-space for a typical choice of interaction constants. All fixed
points are located at the surface of the body $B$.

\begin{figure}[tb]
\includegraphics*[ width=0.7\columnwidth]{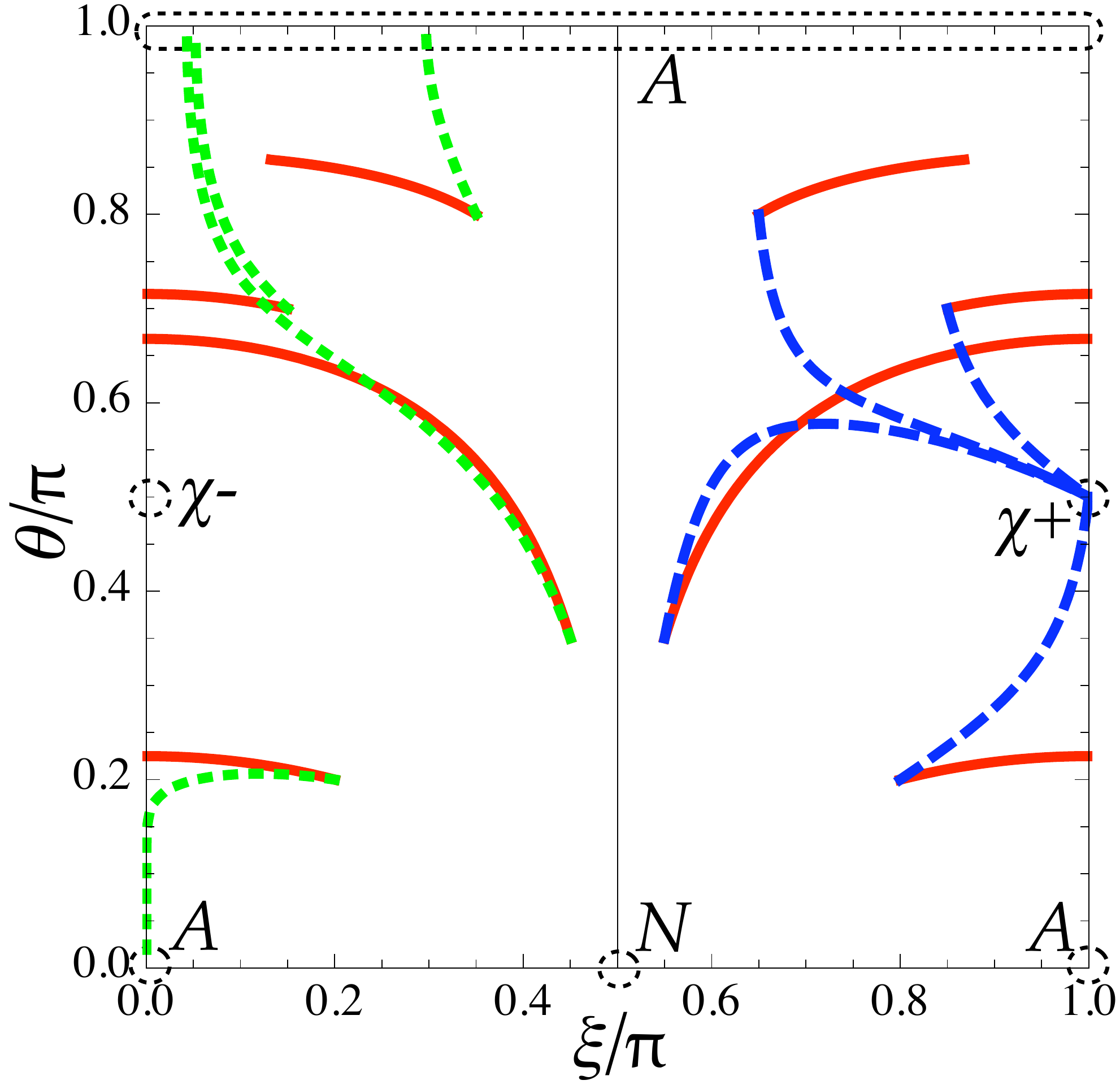}
\caption{(Color online) RG trajectories on the surface $\protect\psi=0$ red
(solid): $g=-0.333$, $g_3=-0.062$, blue (dashed): $g=-0.25$, $g_3=-0.25$,
green (dotted): $g=-0.4$, $g_3=-0.01$. }
\label{fig:RGflows}
\end{figure}

\emph{Lines of fixed points}. As discussed above, the line $g_{3}+g^{2}/2=0$
in the space of interaction constants separates two domains in which fixed
point $\chi^{+}$ (or $\chi^{-}$) or $A$ are stable. For $g$, $g_{3}$ on that
line, in addition to the two fixed points $A,\chi^{\pm}$ we find a whole
line of stable fixed points connecting these two. This means that the RG
flow may stop at any point on this line, depending on the initial
conditions. The fixed point line is parametrized by $a=-1+\frac{1}{2}\tau $, 
$b=1-\frac{3}{2}\tau $, $c=\frac{\sqrt{3}}{2}\tau $, with $\tau\in [ 0,1]$.
Substituting this parametrization into the above RG-equations, we find $%
\frac{d}{d\Lambda}\{a,b,c\}=-\frac{1}{2}(g_{3}+g^{2}/2)\tau(1-\tau)\{1,-3,%
\sqrt{3}\} $, demonstrating that on the line $A$--$\chi^{\pm}$ and for $%
(g_{3}+g^{2}/2)\neq 0$ the flow is directed along the line, towards $A$ or $%
\chi^{\pm}$, depending on whether $-g_{3}<g^{2}/2$ or the opposite. 
When $g_{3}=-g^{2}/2$ the RG flow is directed perpendicular to this line 
with the exponents $g(1\pm \sqrt{1-\tau})$,  i.e.\  
the fixed points on the line are stable. This is demonstrated
in Fig.\ \ref{fig:RGflows} where the RG-trajectories on the surface of body $%
B$ are shown for different $g,g_{3}$ and different initial conditions. We
observe that the blue and green trajectories for interaction strengths in
the regions of interaction parameter space labeled $\chi^{+}$ and $A$
terminate at the respective fixed points (note that the line at $\theta=\pi$
corresponds to a single point), whereas the red lines calculated for
interaction values at the boundary between those two regions are found to
terminate at any point along the lines connecting $A$ and $\chi^{+}$.

There is a second line of fixed points, separating the domain of stability
of the fixed points $M$ ($g>0,g_{3}<0$) and $\chi^{+,-}$ ($g<0,g_{3}<0$). It
is realized for any $g_{3}<0$ at $g=0$. The line of fixed points connects
fixed points $M$ and $\chi^{\pm}$ and is parametrized by $a=-\frac{1}{2}%
\tau^{2}$, $b=-\frac{1}{2}$, $c=\frac{\sqrt{3}}{2}\tau$, with $\tau\in
[-1,1] $. Finally, a trivial line of fixed points exists for any $g_{3}>0$
at $g=0$. It connects the fixed points $A$ and $N$ and corresponds to the
interacting third wire, being detached from the non-interacting wire 1--2 of
arbitrary conductance, $a\in [-1,1]$, $b=1$, $c=0$. Evidently, $a$ should
not be renormalized in this case.

The present weak coupling theory may be extended into the strong coupling
regime, using the approach in \cite{Aristov2011a}, confirming the existence
of lines of fixed points.

\emph{Conclusion}. In this paper we developed a fermionic description of a
Y-junction (1--2 symmetric junction of three TLL-wires) threaded by magnetic
flux, in the regime of weak interaction. In this case the conductance matrix
has three independent components, the conductance associated (i) with
current flowing in wires $1$ and $2$, (ii) with current flowing from wire $3$
into wires $1,2$, (iii) with current in wires $1,2$, driven by a voltage
drop in wire $3$ (Hall conductance). We calculated the scale dependent
contributions to the conductances up to order $g^{2}$ and $g_{3}$ in the
interaction constants. From there we derived three renormalization group
equations for the conductances. These equations have five fixed points
labelled $N,A,M,\chi^{+},\chi^{-}$. When both interactions are repulsive ($%
g>0$ and $g_{3}>0$) only $N$ is stable, which corresponds to the total
separation of the wires. If at least one interaction is attractive ($g$
and/or $g_{3} <0$), we find four regions in $g_{3}$--$g$-space in which $%
N,A,\chi^{\pm},M$ are stable, respectively. At the two border-lines
separating regions of stability of $A,\chi^{\pm}$ (the line $g_{3}+g^{2}/2=0$%
) and of $\chi^{\pm},M$ (the line $g=0$, $g_{3}<0$), respectively, we find a
novel and unexpected situation: at these lines in $g_{3}$--$g$-space not
only the two stable fixed points assigned to the adjacent regions are
stable, but a whole line of fixed points connecting those two fixed points
becomes stable. We have checked that the leading higher order terms do not
destroy the fixed line property (to be published). We conjecture that this
is an incidence of a more general phenomenon: whenever two regions in
interaction constant space with different stable fixed points touch, and the
two fixed points neither vanish nor merge at the contact manifold, a whole
line of stable fixed points connecting the two initial fixed points emerges.

We thank A. Nersesyan, J. Schmalian, V.Yu.\ Kacho\-rovskii, A.P. Dmitriev
and especially I.V. Gornyi and D.G. Polyakov for useful discussions. The
work of D.A. was supported by the German-Israeli Foundation (GIF), the
Dynasty foundation, and a RFBR grant. The work of D.A. and P.W. was
supported by the DFG-Center for Functional Nanostructures at KIT.


\begin{thebibliography}{22}
\expandafter\ifx\csname natexlab\endcsname\relax\def\natexlab#1{#1}\fi
\expandafter\ifx\csname bibnamefont\endcsname\relax
  \def\bibnamefont#1{#1}\fi
\expandafter\ifx\csname bibfnamefont\endcsname\relax
  \def\bibfnamefont#1{#1}\fi
\expandafter\ifx\csname citenamefont\endcsname\relax
  \def\citenamefont#1{#1}\fi
\expandafter\ifx\csname url\endcsname\relax
  \def\url#1{\texttt{#1}}\fi
\expandafter\ifx\csname urlprefix\endcsname\relax\def\urlprefix{URL }\fi
\providecommand{\bibinfo}[2]{#2}
\providecommand{\eprint}[2][]{\url{#2}}

\bibitem[{\citenamefont{Giamarchi}(2003)}]{GiamarchiBook}
\bibinfo{author}{\bibfnamefont{T.}~\bibnamefont{Giamarchi}},
  \emph{\bibinfo{title}{Quantum Physics in One Dimension}}
  (\bibinfo{publisher}{Clarendon Press}, \bibinfo{address}{Oxford},
  \bibinfo{year}{2003}).

\bibitem[{\citenamefont{Kane and Fisher}(1992)}]{Kane1992}
\bibinfo{author}{\bibfnamefont{C.~L.} \bibnamefont{Kane}} \bibnamefont{and}
  \bibinfo{author}{\bibfnamefont{M.~P.~A.} \bibnamefont{Fisher}},
  \bibinfo{journal}{Phys. Rev. B} \textbf{\bibinfo{volume}{46}},
  \bibinfo{pages}{15233} (\bibinfo{year}{1992}).

\bibitem[{\citenamefont{Furusaki and Nagaosa}(1993)}]{Furusaki1993}
\bibinfo{author}{\bibfnamefont{A.}~\bibnamefont{Furusaki}} \bibnamefont{and}
  \bibinfo{author}{\bibfnamefont{N.}~\bibnamefont{Nagaosa}},
  \bibinfo{journal}{Phys. Rev. B} \textbf{\bibinfo{volume}{47}},
  \bibinfo{pages}{4631} (\bibinfo{year}{1993}).

\bibitem[{\citenamefont{Weiss et~al.}(1995)\citenamefont{Weiss, Egger, and
  Sassetti}}]{Weiss1995}
\bibinfo{author}{\bibfnamefont{U.}~\bibnamefont{Weiss}},
  \bibinfo{author}{\bibfnamefont{R.}~\bibnamefont{Egger}}, \bibnamefont{and}
  \bibinfo{author}{\bibfnamefont{M.}~\bibnamefont{Sassetti}},
  \bibinfo{journal}{Phys. Rev. B} \textbf{\bibinfo{volume}{52}},
  \bibinfo{pages}{16707} (\bibinfo{year}{1995}).

\bibitem[{\citenamefont{Fendley et~al.}(1995)\citenamefont{Fendley, Ludwig, and
  Saleur}}]{Fendley1995}
\bibinfo{author}{\bibfnamefont{P.}~\bibnamefont{Fendley}},
  \bibinfo{author}{\bibfnamefont{A.~W.~W.} \bibnamefont{Ludwig}},
  \bibnamefont{and} \bibinfo{author}{\bibfnamefont{H.}~\bibnamefont{Saleur}},
  \bibinfo{journal}{Phys. Rev. B} \textbf{\bibinfo{volume}{52}},
  \bibinfo{pages}{8934} (\bibinfo{year}{1995}).

\bibitem[{\citenamefont{Yue et~al.}(1994)\citenamefont{Yue, Glazman, and
  Matveev}}]{Yue1994}
\bibinfo{author}{\bibfnamefont{D.}~\bibnamefont{Yue}},
  \bibinfo{author}{\bibfnamefont{L.~I.} \bibnamefont{Glazman}},
  \bibnamefont{and} \bibinfo{author}{\bibfnamefont{K.~A.}
  \bibnamefont{Matveev}}, \bibinfo{journal}{Phys. Rev. B}
  \textbf{\bibinfo{volume}{49}}, \bibinfo{pages}{1966} (\bibinfo{year}{1994}).

\bibitem[{\citenamefont{Polyakov and Gornyi}(2003)}]{Polyakov2003}
\bibinfo{author}{\bibfnamefont{D.~G.} \bibnamefont{Polyakov}} \bibnamefont{and}
  \bibinfo{author}{\bibfnamefont{I.~V.} \bibnamefont{Gornyi}},
  \bibinfo{journal}{Phys. Rev. B} \textbf{\bibinfo{volume}{68}},
  \bibinfo{pages}{035421} (\bibinfo{year}{2003}).

\bibitem[{\citenamefont{Das et~al.}(2004)\citenamefont{Das, Rao, and
  Sen}}]{Das2004}
\bibinfo{author}{\bibfnamefont{S.}~\bibnamefont{Das}},
  \bibinfo{author}{\bibfnamefont{S.}~\bibnamefont{Rao}}, \bibnamefont{and}
  \bibinfo{author}{\bibfnamefont{D.}~\bibnamefont{Sen}},
  \bibinfo{journal}{Phys. Rev. B} \textbf{\bibinfo{volume}{70}},
  \bibinfo{pages}{085318} (\bibinfo{year}{2004}).

\bibitem[{\citenamefont{Lal et~al.}(2002)\citenamefont{Lal, Rao, and
  Sen}}]{Lal2002}
\bibinfo{author}{\bibfnamefont{S.}~\bibnamefont{Lal}},
  \bibinfo{author}{\bibfnamefont{S.}~\bibnamefont{Rao}}, \bibnamefont{and}
  \bibinfo{author}{\bibfnamefont{D.}~\bibnamefont{Sen}},
  \bibinfo{journal}{Phys. Rev. B} \textbf{\bibinfo{volume}{66}},
  \bibinfo{pages}{165327} (\bibinfo{year}{2002}).

\bibitem[{\citenamefont{Aristov and W\"{o}lfle}(2009)}]{Aristov2009}
\bibinfo{author}{\bibfnamefont{D.~N.} \bibnamefont{Aristov}} \bibnamefont{and}
  \bibinfo{author}{\bibfnamefont{P.}~\bibnamefont{W\"{o}lfle}},
  \bibinfo{journal}{Phys. Rev. B} \textbf{\bibinfo{volume}{80}},
  \bibinfo{eid}{045109}   (\bibinfo{year}{2009}).

\bibitem[{\citenamefont{Chamon et~al.}(2003)\citenamefont{Chamon, Oshikawa, and
  Affleck}}]{Chamon2003}
\bibinfo{author}{\bibfnamefont{C.}~\bibnamefont{Chamon}},
  \bibinfo{author}{\bibfnamefont{M.}~\bibnamefont{Oshikawa}}, \bibnamefont{and}
  \bibinfo{author}{\bibfnamefont{I.}~\bibnamefont{Affleck}},
  \bibinfo{journal}{Phys. Rev. Lett.} \textbf{\bibinfo{volume}{91}},
  \bibinfo{pages}{206403} (\bibinfo{year}{2003}).

\bibitem[{\citenamefont{Oshikawa et~al.}(2006)\citenamefont{Oshikawa, Chamon,
  and Affleck}}]{Oshikawa2006}
\bibinfo{author}{\bibfnamefont{M.}~\bibnamefont{Oshikawa}},
  \bibinfo{author}{\bibfnamefont{C.}~\bibnamefont{Chamon}}, \bibnamefont{and}
  \bibinfo{author}{\bibfnamefont{I.}~\bibnamefont{Affleck}},
  \bibinfo{journal}{J. Stat. Mech.} \textbf{\bibinfo{volume}{2006}},
  \bibinfo{pages}{P02008} (\bibinfo{year}{2006}).

\bibitem[{\citenamefont{Barnab\'e-Th\'eriault
  et~al.}(2005)\citenamefont{Barnab\'e-Th\'eriault, Sedeki, Meden, and
  Sch\"onhammer}}]{Barnabe2005}
\bibinfo{author}{\bibfnamefont{X.}~\bibnamefont{Barnab\'e-Th\'eriault}},
  \bibinfo{author}{\bibfnamefont{A.}~\bibnamefont{Sedeki}},
  \bibinfo{author}{\bibfnamefont{V.}~\bibnamefont{Meden}}, \bibnamefont{and}
  \bibinfo{author}{\bibfnamefont{K.}~\bibnamefont{Sch\"onhammer}},
  \bibinfo{journal}{Phys. Rev. Lett.} \textbf{\bibinfo{volume}{94}},
  \bibinfo{pages}{136405} (\bibinfo{year}{2005}).

\bibitem[{\citenamefont{Bellazzini et~al.}(2007)\citenamefont{Bellazzini,
  Mintchev, and Sorba}}]{Bellazzini2007}
\bibinfo{author}{\bibfnamefont{B.}~\bibnamefont{Bellazzini}},
  \bibinfo{author}{\bibfnamefont{M.}~\bibnamefont{Mintchev}}, \bibnamefont{and}
  \bibinfo{author}{\bibfnamefont{P.}~\bibnamefont{Sorba}},
  \bibinfo{journal}{Journal of Physics A: Mathematical and Theoretical}
  \textbf{\bibinfo{volume}{40}}, \bibinfo{pages}{2485} (\bibinfo{year}{2007}).

\bibitem[{\citenamefont{Hou and Chamon}(2008)}]{Hou2008}
\bibinfo{author}{\bibfnamefont{C.-Y.} \bibnamefont{Hou}} \bibnamefont{and}
  \bibinfo{author}{\bibfnamefont{C.}~\bibnamefont{Chamon}},
  \bibinfo{journal}{Phys. Rev. B} \textbf{\bibinfo{volume}{77}},
  \bibinfo{pages}{155422} (\bibinfo{year}{2008}).

\bibitem[{\citenamefont{Hou et~al.}(2009)\citenamefont{Hou, Kim, and
  Chamon}}]{Hou2009}
\bibinfo{author}{\bibfnamefont{C.-Y.} \bibnamefont{Hou}},
  \bibinfo{author}{\bibfnamefont{E.-A.} \bibnamefont{Kim}}, \bibnamefont{and}
  \bibinfo{author}{\bibfnamefont{C.}~\bibnamefont{Chamon}},
  \bibinfo{journal}{Phys. Rev. Lett.} \textbf{\bibinfo{volume}{102}},
  \bibinfo{pages}{076602} (\bibinfo{year}{2009}).

\bibitem[{\citenamefont{Das and Rao}(2011)}]{Das2011}
\bibinfo{author}{\bibfnamefont{S.}~\bibnamefont{Das}} \bibnamefont{and}
  \bibinfo{author}{\bibfnamefont{S.}~\bibnamefont{Rao}},
  \bibinfo{journal}{Phys. Rev. Lett.} \textbf{\bibinfo{volume}{106}},
  \bibinfo{pages}{236403} (\bibinfo{year}{2011}).

\bibitem[{\citenamefont{Steinberg et~al.}(2008)\citenamefont{Steinberg, Barak,
  Yacoby, Pfeiffer, West, Halperin, and Le~Hur}}]{Steinberg2008}
\bibinfo{author}{\bibfnamefont{H.}~\bibnamefont{Steinberg}},
  \bibinfo{author}{\bibfnamefont{G.}~\bibnamefont{Barak}},
  \bibinfo{author}{\bibfnamefont{A.}~\bibnamefont{Yacoby}},
  \bibinfo{author}{\bibfnamefont{L.~N.} \bibnamefont{Pfeiffer}},
  \bibinfo{author}{\bibfnamefont{K.~W.} \bibnamefont{West}},
  \bibinfo{author}{\bibfnamefont{B.~I.} \bibnamefont{Halperin}},
  \bibnamefont{and} \bibinfo{author}{\bibfnamefont{K.}~\bibnamefont{Le~Hur}},
  \bibinfo{journal}{Nature Physics} \textbf{\bibinfo{volume}{4}},
  \bibinfo{pages}{116} (\bibinfo{year}{2008}).

\bibitem[{\citenamefont{Meden et~al.}(2008)\citenamefont{Meden, Andergassen,
  Enss, Schoeller, and Sch\"onhammer}}]{Meden2008}
\bibinfo{author}{\bibfnamefont{V.}~\bibnamefont{Meden}},
  \bibinfo{author}{\bibfnamefont{S.}~\bibnamefont{Andergassen}},
  \bibinfo{author}{\bibfnamefont{T.}~\bibnamefont{Enss}},
  \bibinfo{author}{\bibfnamefont{H.}~\bibnamefont{Schoeller}},
  \bibnamefont{and}
  \bibinfo{author}{\bibfnamefont{K.}~\bibnamefont{Sch\"onhammer}},
  \bibinfo{journal}{New Journal of Physics} \textbf{\bibinfo{volume}{10}},
  \bibinfo{pages}{045012} (\bibinfo{year}{2008}).

\bibitem[{\citenamefont{Aristov et~al.}(2010)\citenamefont{Aristov, Dmitriev,
  Gornyi, Kachorovskii, Polyakov, and W\"olfle}}]{Aristov2010}
\bibinfo{author}{\bibfnamefont{D.~N.} \bibnamefont{Aristov}},
  \bibinfo{author}{\bibfnamefont{A.~P.} \bibnamefont{Dmitriev}},
  \bibinfo{author}{\bibfnamefont{I.~V.} \bibnamefont{Gornyi}},
  \bibinfo{author}{\bibfnamefont{V.~Y.} \bibnamefont{Kachorovskii}},
  \bibinfo{author}{\bibfnamefont{D.~G.} \bibnamefont{Polyakov}},
  \bibnamefont{and} \bibinfo{author}{\bibfnamefont{P.}~\bibnamefont{W\"olfle}},
  \bibinfo{journal}{Phys. Rev. Lett.} \textbf{\bibinfo{volume}{105}},
  \bibinfo{pages}{266404} (\bibinfo{year}{2010}).

\bibitem[{\citenamefont{Aristov and W\"olfle}(2011)}]{Aristov2011a}
\bibinfo{author}{\bibfnamefont{D.~N.} \bibnamefont{Aristov}} \bibnamefont{and}
  \bibinfo{author}{\bibfnamefont{P.}~\bibnamefont{W\"olfle}},
  \bibinfo{journal}{Phys. Rev. B} \textbf{\bibinfo{volume}{84}},
  \bibinfo{pages}{155426} (\bibinfo{year}{2011}).
  
  

\bibitem{footnote} ``Lines of fixed points'' discussed in \cite{Barnabe2005}
correspond to one point ($N$) in terms of conductance.

\bibitem[{\citenamefont{Aristov}(2011)}]{Aristov2011}
\bibinfo{author}{\bibfnamefont{D.~N.} \bibnamefont{Aristov}},
  \bibinfo{journal}{Phys. Rev. B} \textbf{\bibinfo{volume}{83}},
  \bibinfo{pages}{115446} (\bibinfo{year}{2011}).


\bibitem[{\citenamefont{Rahmani et~al.}(2010)\citenamefont{Rahmani, Hou,
  Feiguin, Chamon, and Affleck}}]{Rahmani2010}
\bibinfo{author}{\bibfnamefont{A.}~\bibnamefont{Rahmani}},
  \bibinfo{author}{\bibfnamefont{C.-Y.} \bibnamefont{Hou}},
  \bibinfo{author}{\bibfnamefont{A.}~\bibnamefont{Feiguin}},
  \bibinfo{author}{\bibfnamefont{C.}~\bibnamefont{Chamon}}, \bibnamefont{and}
  \bibinfo{author}{\bibfnamefont{I.}~\bibnamefont{Affleck}},
  \bibinfo{journal}{Phys. Rev. Lett.} \textbf{\bibinfo{volume}{105}},
  \bibinfo{pages}{226803} (\bibinfo{year}{2010}).

\end{thebibliography}

\end{document}